\documentclass[12pt,preprint]{aastex}

\newcommand{\arcs}{\mbox{\ensuremath{^{\prime\prime}}}}
\newcommand{\arcm}{\mbox{\ensuremath{^{\prime}}}}

\usepackage{lscape,natbib}
\begin{document}

\title{Photoevaporating Proplyd-like objects in Cygnus OB2}

\author{Nicholas J. Wright$^1$, Jeremy J. Drake$^1$, Janet E. Drew$^2$, Mario G. Guarcello$^1$,\\ Robert A. Gutermuth$^{3}$, Joseph L. Hora$^1$, and Kathleen E. Kraemer$^4$}
\affil{
$^1$ Harvard-Smithsonian Center for Astrophysics, 60 Garden Street, Cambridge, MA~02138, USA\\
$^2$ Centre for Astronomy Research, Science and Technology Research Institute, University of Hertfordshire, Hatfield AL10~9AB, UK\\
$^3$ Dept. of Astronomy, University of Massachusetts, Amherst, MA, USA\\
$^4$ Boston College, Institute for Scientific Research, 140 Commonwealth Avenue, Chestnut Hill, MA 02467, USA\\
}
\email{nwright@cfa.harvard.edu}

\begin{abstract}

We report the discovery of ten proplyd-like objects in the vicinity of the massive OB association Cygnus~OB2. They were discovered in IPHAS H$\alpha$ images and are clearly resolved in broad-band HST/ACS, near-IR and {\it Spitzer} mid-IR images. All exhibit the familiar tadpole shape seen in photoevaporating objects such as the Orion proplyds, with a bright ionization front at the head facing the central cluster of massive stars, and a tail stretching in the opposite direction. Many also show secondary ionization fronts, complex tail morphologies or multiple heads. We consider the evidence that these are either proplyds or `evaporating gaseous globules' (EGGs) left over from a fragmenting molecular cloud, but find that neither scenario fully explains the observations. Typical sizes are 50,000--100,000~AU, larger than the Orion proplyds, but in agreement with the theoretical scaling of proplyd size with distance from the ionizing source. These objects are located at projected separations of $\sim$6--14~pc from the OB association, compared to $\sim$0.1~pc for the Orion proplyds, but are clearly being photoionized by the $\sim$65 O-type stars in Cyg~OB2. Central star candidates are identified in near- and mid-IR images, supporting the proplyd scenario, though their large sizes and notable asymmetries is more consistent with the EGG scenario. A third possibility is therefore considered, that these are a unique class of photoevaporating partially-embedded young stellar objects that have survived the destruction of their natal molecular cloud. This has implications for the properties of stars that form in the vicinity of massive stars.

\end{abstract}


\section{Introduction}

How the star formation process differs in the harsher environments in the Milky Way is of particular interest when considering star formation products such as the initial mass function, the binary fraction, and the fraction of stars that may be home to planetary systems.  In recent years evidence has grown that circumstellar disks around young stars are eroded by UV radiation from massive stars \citep[e.g.,][]{john98,guar07}. Since these disks are believed to be the origin of planetary systems such as our own, and the majority of stars that form in our Galaxy do so in regions containing massive stars, this is of particular importance for a global understanding of star formation.

One of the most significant advances in our understanding of the evolution of protoplanetary disks was the discovery of ``proplyds'' in the Orion Nebula \citep{laqu79,chur87,odel94}, young stellar objects (YSOs) whose circumstellar material renders them visible either through direct emission or silhouetted against a background H{\sc ii} region. The proplyds seen in emission have distinctive tadpole shapes with tails pointing away from $\theta^1$~Ori~C \citep{odel93,odel96} that are well explained by theoretical models of UV photoevaporation \citep{john98}.

Proplyds provide illustrative evidence of the evaporation of circumstellar disks in harsh environments.  It is therefore useful to identify them in other systems where photoevaporation might be more widespread.  Searches in more massive clusters are hindered by the typically greater distances to them, often leading to the mis-identification of `evaporating gaseous globules' (EGGs) - fragments of an evaporating molecular cloud - as proplyds \citep[e.g.,][]{smit03,dema06,smit10}. Candidate proplyds have so far been identified in NGC~2244, IC~1396, NGC~2264 \citep[all][]{balo06}, W5 \citep{koen08}, the Trifid Nebula \citep{yuse05}, the Lagoon Nebula \citep{stec98}, and the Carina star-forming region \citep{smit03}, but the best candidates outside of Orion are those in NGC~3603 \citep{bran00}, although these are significantly larger than the Orion proplyds. Many of these regions contain only a single candidate proplyd and never has a true family of proplyds similar to those in the Orion Nebula been found in another region.

Cygnus~OB2 is the most massive young association within 2~kpc of the Sun.  It lies within the Cygnus~X giant molecular cloud where considerable star formation is ongoing \citep[e.g.,][]{schn06}.  The association contains $>$65 O-type stars \citep{mass91,come02,hans03,kimi07} and therefore provides an excellent sample of the influence of massive stars on the evolution of circumstellar disks. A recent X-ray survey by \citet{wrig09a} identified $\sim$1500 lower-mass YSOs, while \citet{wrig10a} found evidence for an age spread among them from 3--5~Myr.  This is consistent with evidence for O stars as young as 1~Myr \citep{hans03}, and an older population with an age of 5--7~Myr \citep{drew08}. The central cluster has cleared away the majority of molecular gas in the vicinity of the association \citep{schn06}, but recent observations have revealed sites of ongoing star formation around the periphery \citep[e.g.,][]{vink08}.

In this paper, we present the discovery of 10 objects in the vicinity of Cygnus OB2 that appear remarkably similar to the Orion proplyds. In Section~2 we present the observations used to identify and characterize them, in Section~3 we discuss their physical properties and consider their true nature, and finally in Section~4 we discuss the implications of this finding for our understanding of the propagation of star formation in Cygnus~OB2.  We adopt a distance of 1.40~kpc for Cyg~OB2 \citep{rygl11}, whereby $1\arcm = 0.41$~pc.

\section{Observations}

The objects presented here were discovered through visual inspection of INT (Isaac Newton Telescope) Photometric H$\alpha$ Survey \citep[IPHAS,][]{drew05} H$\alpha$ images of Cygnus~OB2. Here we present the observations that led to their discovery and complementary data at other wavelengths.  We enumerate the objects and adopt the IPHAS naming system based on their right ascension and declination.  One of these objects, \#5, was identified as an H$\alpha$ emission source by \citet{viir09} and labelled IPHASX~J203311.5+404141 during an automated search of IPHAS images for compact planetary nebulae.

\subsection{IPHAS H$\alpha$ Observations}

\begin{figure*}
\includegraphics[width=470pt]{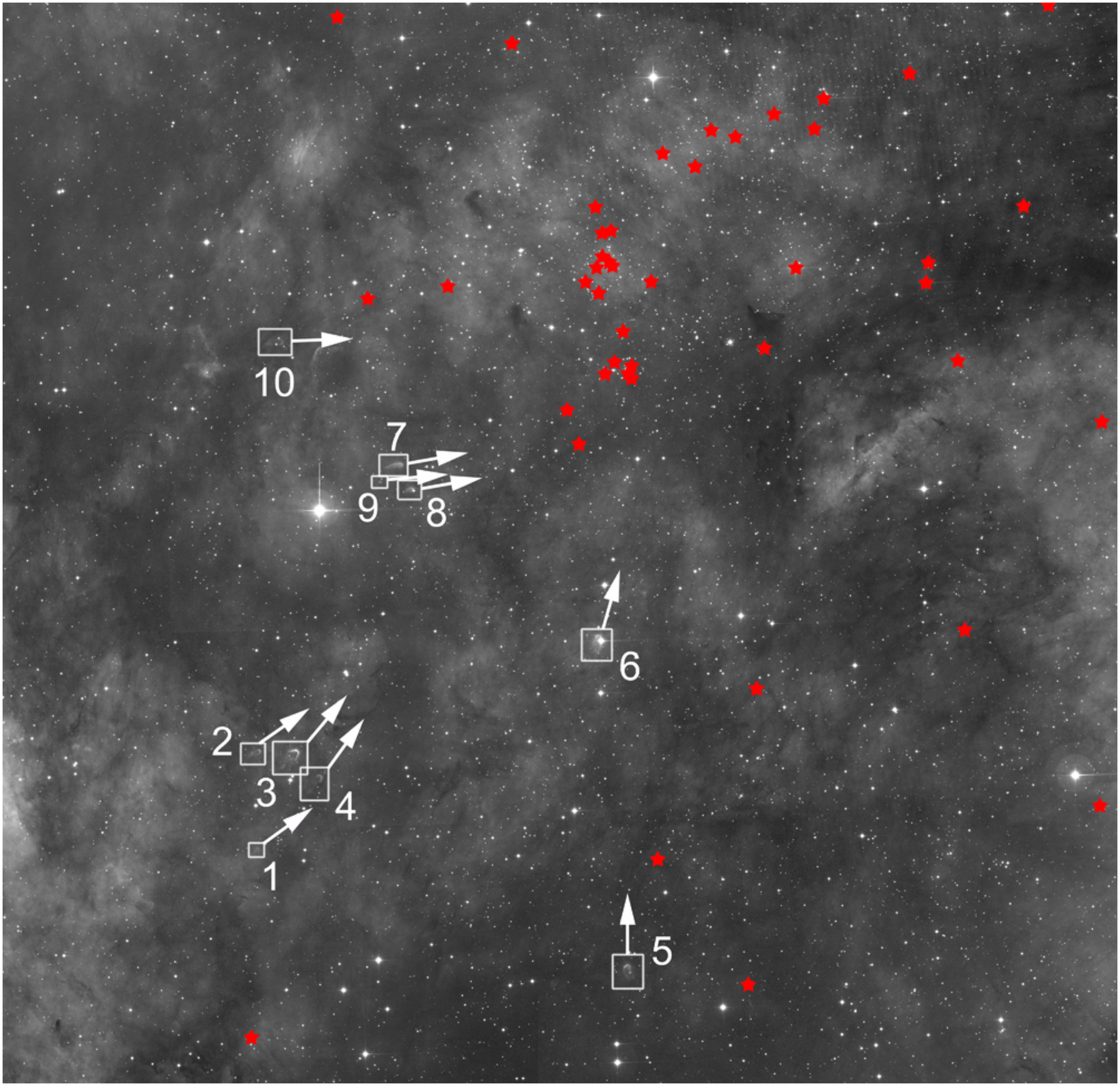}
\caption{IPHAS H$\alpha$ image of the Cygnus~OB2 region displayed using a logarithmic intensity scale. The image is $1^\circ \times 1^\circ$ ($\sim 24 \times 24$~pc at 1.40~kpc) and centered on (R.A., Dec) = (20:33:20, +41:05:00) with north up and east to the left. The locations of the ten objects are indicated and numbered, with arrows showing the direction of their semi-major axes and ionization fronts. The positions of all known O-type stars are shown as red stars. DR~15 is approximately 20\arcm\ south of this image.}
\label{cygnus_wide}
\end{figure*}

IPHAS is a survey of the northern Galactic plane in broad-band Sloan $r'$, $i'$, and narrow-band H$\alpha$ filters using the INT.  Tiling of the 0.25~deg$^2$ Wide Field Camera field-of-view leads to almost uninterrupted survey coverage across the Galactic plane at high spatial resolution (1 pixel = 0.333\arcs). The use of a narrow-band H$\alpha$ filter ($\lambda_c = 6568$ \AA, FWHM = 95 \AA) picks out extended emission-line nebulae and point sources \citep[e.g.,][]{ware06,with08,sabi10}. A $1.5 \times 1.5$ degree H$\alpha$ mosaic of the Cygnus region was produced using {\sc montage}, part of which is shown in Figure~\ref{cygnus_wide} with the positions of the ten objects discussed in this work indicated. This image was searched for other structures similar to these, but none were found down to the limiting resolution of the image. Individual images of the objects in the H$\alpha$ filter are shown in Figures~2--4. The $r'$ and $i'$ broad-band images provide no further information on these diffuse structures and are omitted.

\subsection{UKIDSS near-IR Observations}

Near-IR observations were taken from the UKIRT Infrared Deep Sky Survey \citep[UKIDSS,][pixel size 0.4\arcs]{lawr07} Galactic Plane Survey \citep[GPS,][]{luca08}. The $J$ and $H$ band images are broadly similar to the deeper $K$-band image shown in Figures~2--4.

\subsection{{\it Spitzer} mid-IR Observations}

The {\it Spitzer} Cygnus-X Legacy Survey \citep{hora11} imaged Cygnus~X with the Infrared Array Camera \citep[IRAC;][]{fazi04} in four bands (3.6, 4.5, 5.8, and 8.0 $\mu$m) and the Multiband Imaging Photometer for {\it Spitzer} \citep[MIPS;][]{riek04} in the 24~$\mu$m band.  Figures~2--4 show the objects in the IRAC 8.0~$\mu$m (pixel size of 1.2\arcs), and MIPS 24~$\mu$m (pixel size of 2.55\arcs) bands.

\subsection{HST/ACS Observations}

Object \#7, also known as IRAS~20324+4057, was observed by the Hubble Space Telescope (HST) on 2006 July 22 with the Advanced Camera for Surveys (ACS, 0.05\arcs / pixel).  Observations used the broad-band filters F606W ($\lambda_c = 5907$ \AA, $\Delta \lambda = 2342$ \AA, approximately $V$+$R$) and F814W ($\lambda_c = 8333$ \AA, $\Delta \lambda = 2511$ \AA, approximately a broad $I$ filter).  Two observations were made in each filter, for a total exposure of 694s per filter.  A color image of \#7 compiled from these observations is shown in Figure~\ref{images3}.

\subsection{Previous identifications}

Some of these objects have previous identifications, either through their extended emission or their central stars, however none have previously been identified as proplyds or EGGs. Some were detected in the infrared, the sub-mm, or radio (see Table~\ref{proplyds}).  Objects 3, 5, and 7 were identified as possible OB stars by \citet{come02} based on near-IR photometry and low-resolution spectroscopy, suggesting that the central stars of these objects may be more massive than the typical proplyds seen in Orion.  They found that \#3 exhibits possible H$_2$ and CO emission that could originate from a massive circumstellar disk, as the mid-IR images corroborate, while \#7 shows Br$\gamma$ in emission, most likely originating from the nebulosity.  Object \#7 was also imaged by \citet{pere07b} during a search for post-AGB candidates and was identified as an H{\sc ii} region with a bow shock. Spectra revealed nebular emission lines in a ratio that implied photoionization consistent with ionization by massive stars, rather than shock-excitation.

\section{Discussion}
\label{discuss}

We now discuss the structure and morphology of the objects and consider the evidence that they are true proplyds like those seen in Orion, or if they are EGGs like those observed by \citet{mcca02} and \citet{dema06}. We note that, unlike many of the Orion proplyds, these objects are all seen in emission and not in silhouette \citep[e.g.,][]{mcca96}. The silhouettes in Orion are  the actual disks themselves, which would not be resolvable at the greater distance of Cygnus~OB2 with the observations available. The positions and sizes of these objects are summarized in Table~\ref{proplyds}. 

\subsection{Morphologies}

All the structures in Figures~2--4 are clearly tadpole shaped with extended tails pointing away from Cyg~OB2. They are rim-brightened on the side facing the OB association clearly indicating that they are being photoionized by the massive stars in Cyg~OB2. Some show very similar morphologies to Orion proplyds.  For example, \#4 shares the `safety pin' shape of d177-341 \citep{ball00}, where the thin sides of the proplyd appear to join up at the end of its tail in a small clump.  However, there are minor differences between these proplyds and those in Orion.  Objects 2, 5, \& 6 show distinctly non-axisymmetric shapes with many small clumps. The HST image of \#7 in Figure~\ref{images3} also reveals an interesting non-axisymmetry: the ionization brightness varies between the northern and southern sides of its tail.  A foreground extinction gradient might cause this, although the required gradient angular size of $\sim$10$^{\prime\prime}$ renders this unlikely.

These asymmetries may suggest that these are not proplyds, since the Orion proplyds are clearly symmetric. However, proplyd morphology is determined by the ionizing flux \citep{john98}, and here Cygnus~OB2 differs significantly from the Orion Nebula or NGC~3603. The massive stars in Cyg~OB2 are distributed over a wider area (Figure~1), and not centrally-concentrated like those in other regions, so the ionizing flux emanates from a broader area. \citet{john98} showed that the ratio of far-UV to extreme-UV flux is important in determining the properties of the resulting photo-evaporative flow and photoionized morphology.  Since this ratio varies across the O-type spectral class, the positions of different O-type stars across Cygnus~OB2 will result in variation of the flow in different directions and could explain the observed asymmetries.

\subsection{Sizes}

The lengths of these objects range from 18,000 to 113,000~AU (Table~\ref{proplyds}), making them larger than both the Orion proplyds \citep[40--400~AU,][]{odel98}, and the hitherto largest-known `giant proplyds' in NGC~3603 \citep[$\sim$15,000--21,000~AU,][]{bran00}.  The absence of smaller objects is a selection effect due to the greater distance of Cyg~OB2 and the ground-based imaging from which they were discovered. The large sizes of these objects, orders of magnitude larger than the Orion proplyds, suggests that they could be EGGs, which occupy a large range of sizes.

Despite this the large sizes of these objects can actually be explained by current models of proplyd photoevaporation \citep{john98}. Proplyd size depends on how far the ionization front caused by stellar Lyman continuum photons can penetrate into the outflowing photo-dissociation region formed from evaporating circumstellar material (heated by Balmer continuum photons).  The size of the proplyd scales inversely with the ionizing extreme UV flux, and therefore should scale with the distance to the source of ionizing radiation, as seen in the Orion Nebula \citep{odel98}.  

While the O stars in Cyg~OB2 are spread over a large area, we can approximate the center of the association to be mid-way between the two main groups of massive stars, the northern clump containing Cyg~OB2 \#8 and the southern clump dominated by Cyg~OB2 \#22 \citep[see e.g.,][]{negu08}.  These proplyd-like objects are at projected distances from this center of $\sim$14--34$^\prime$, or $\sim$6--14~pc (see Table~\ref{proplyds}).  We find no correlation between this projected distance and the length of the object, suggesting that both these objects and the O stars are sufficiently distributed along the line of sight that their projected separations are misleading.  However, the typical sizes of the Orion and NGC~3603 proplyds and these objects ($\sim$100, $\sim$20,000, and $\sim$70,000~AU respectively) do scale with their projected distances from the source of ionization ($< 0.1$~pc, 1--2~pc, and $\sim$6--14~pc, respectively), supporting a broader correlation.

\subsection{Candidate central stars and the true nature of these objects}

Proplyds, by definition, include an embedded YSO with a protoplanetary disk, while fragments of molecular cloud material may not. In their deep near-IR study of EGGs in the Eagle Nebula, \citet{mcca02} could only identify protostars in 15\% of the globules, at extinctions up to $A_V \sim 27$. This suggests that either most EGGs do not contain central stars, or if they do are so heavily embedded as to escape detection.

To assess the fraction of these objects that contain central stars we searched for point sources in the UKIDSS near-IR images, finding good candidates in 7 of the 10 objects. This fraction is significantly higher than that found by \citet{mcca02}, despite the fact that our observations are $\sim$3 magnitudes shallower than theirs (the two regions are at similar distances). All these sources are also detected in {\it Spitzer} mid-IR observations, suggesting these sources have circumstellar material. The remaining 3 sources appear more complex; \#1 shows mid-IR emission, but is faint and not clearly associated with a central star candidate; \#10 shows only a very faint near-IR point source and no mid-IR source; and \#6 has no good candidate near- or mid-IR source. The detected central source fraction of 70\% appears more consistent with the proplyd scenario than with the EGG scenario.

Many of these point sources are particularly bright at 24~$\mu$m, suggesting a significant mass of cool dust that would be more consistent with embedded Class~I sources as opposed to disk-bearing Class~II sources. This fact, coupled with the failure of either the proplyd or EGG scenarios to fully explain all the observations, leads us to consider a third scenario. These objects may represent a more advanced stage of EGG evolution, whereby empty EGGs have fully evaporated but those with central stars, which may also be the densest and most massive, have survived the photoevaporation process. This scenario requires further verification but explains many of the observations that the other scenarios do not. If correct it raises new questions such as how the photoevaporation of star-forming EGGs influences the final mass of the embedded stars and how this affects the initial mass function in the vicinity of massive clusters.

\section{The propagation of star formation in Cygnus}

Regardless of their true nature the presence of clear central star candidates with circumstellar material implies that these objects are very young. If they are proplyds it is well known that irradiated protoplanetary disks are thought to be evaporated in $<$1~Myr \citep{chur87,odel08}, whereas if these are some sort of star-forming EGG they are likely to be even younger. These objects must therefore represent a younger and more-recent population of star formation in the Cygnus region than the slightly older Cyg~OB2 association itself. The wide-field view in Figure~\ref{cygnus_wide} shows the location of the objects relative to the Cyg~OB2 association. The objects are located to the south-east of the main core of the association, approximately half-way between Cyg~OB2 and the recently discovered cluster in the H{\sc ii} region DR~15 \citep{vink08}, supporting propagation of star formation in a southerly direction. The formation of these objects and the DR~15 cluster may therefore have been triggered by stellar winds or UV radiation from the massive stars in Cyg~OB2 or they may have formed anyway and are only now being influenced by Cyg~OB2. Differentiating between these two scenarios is important for understand the propagation of star formation in massive clusters and associations such as this.

\section{Summary}

We have presented multi-wavelength images of 10 proplyd-like objects in the vicinity of the nearby massive association Cygnus~OB2. Their tadpole-like shapes, clearly visible in IPHAS H$\alpha$ images, are oriented towards the central OB association. We have considered the arguments that these are either true proplyds like those in Orion, or photoionized fragments of molecular cloud material known as EGGs. These objects are notably different from the Orion proplyds, in terms of their larger sizes, asymmetries, and large distances from the source of ionization, but we argue that these factors do not necessarily disprove that they are proplyds and can all be explained under the current model of disk photoevaporation. We show that their large sizes are in agreement with the predicted scaling of proplyd size with distance from the ionizing source, and argue that their asymmetries can be partly explained by the extended distribution of O stars across Cyg~OB2 and therefore the complex UV flux distribution. We identify embedded central stars in 70\% of the objects using near- and mid-IR images, a significantly larger fraction than observed in EGGs in other star forming regions and closer to the 100\% expected for proplyds. We conclude that neither scenario adequately explains the observed properties of these objects and therefore consider a third possibility that these are an intermediate class of object representing the subset of star-forming EGGs that have survived the photoevaporation of the molecular cloud. If this is the case, and these objects are still accreting from their natal molecular cloud, it raises the question of whether photoevaporation from massive stars could influence the initial mass function in regions such as this.

A future paper will present a complete analysis of the photometric properties of the central stars and the long-wavelength emission from circumstellar material, combined with an analysis of the UV radiation field from Cygnus~OB2. Further observations may be necessary to determine the true nature of these objects, such as spectroscopy of the central stars and the ionized gas associated with them, or CO observations to ascertain the mass of the gas reservoir.

\acknowledgments

We thank Geert Barentsen, Robert Greimel, Roberto Raddi, and Jorick Vink for constructive discussions and thank the anonymous referee for a prompt report. This research used data obtained at the INT by the IPHAS project and processed by the Cambridge Astronomical Survey Unit.  We acknowledge data products from the {\it Spitzer} Space Telescope, UKIDSS, and the NASA/ESA Hubble Space Telescope, and {\sc montage}, funded by NASA.  JJD was supported by NASA contract NAS8-39073 to the {\it Chandra} X-ray Center (CXC).

%\bibliography{/Users/nwright/Documents/Work/tex_papers/bibliography.bib}
%\bibliographystyle{apj}
% papers with more than 5 authors truncated to first three with {et al}

\begin{figure*}
\includegraphics[scale=1.0, trim=0 -5 0 -50, clip=true]{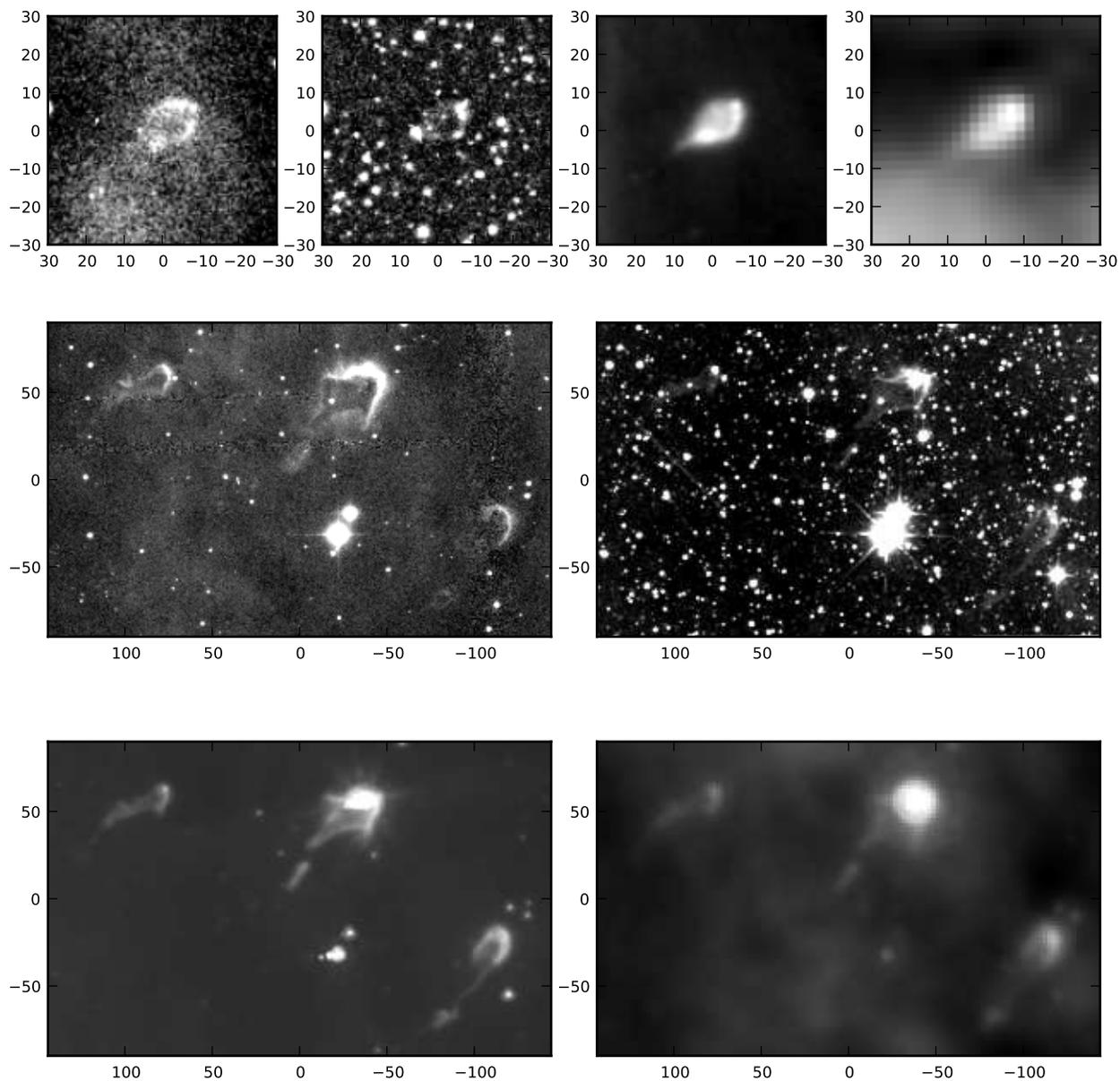}
\caption{Images of objects \#1--4 at different wavelengths. {\it Top:} Object \#1, from left to right: IPHAS H$\alpha$, UKIDSS K-band, IRAC~8~$\mu$m, MIPS~24~$\mu$m. {\it Middle:} Objects \#2--4 in IPHAS H$\alpha$ (left) and UKIDSS K-band (right). {\it Bottom:} Objects \#2--4 in IRAC 8~$\mu$m (left) and MIPS 24~$\mu$m. All images are shown with north up and east to the left and with axes in arcseconds offset from the center of each image. At a distance of 1.40~kpc, $10\arcs \sim 1.4 \times 10^4 \mathrm{AU} \sim 0.07$~pc.}
\label{images1}
\end{figure*}

\clearpage

\begin{figure*}
\includegraphics[scale=1.0, trim=-10 -5 -10 -225, clip=true]{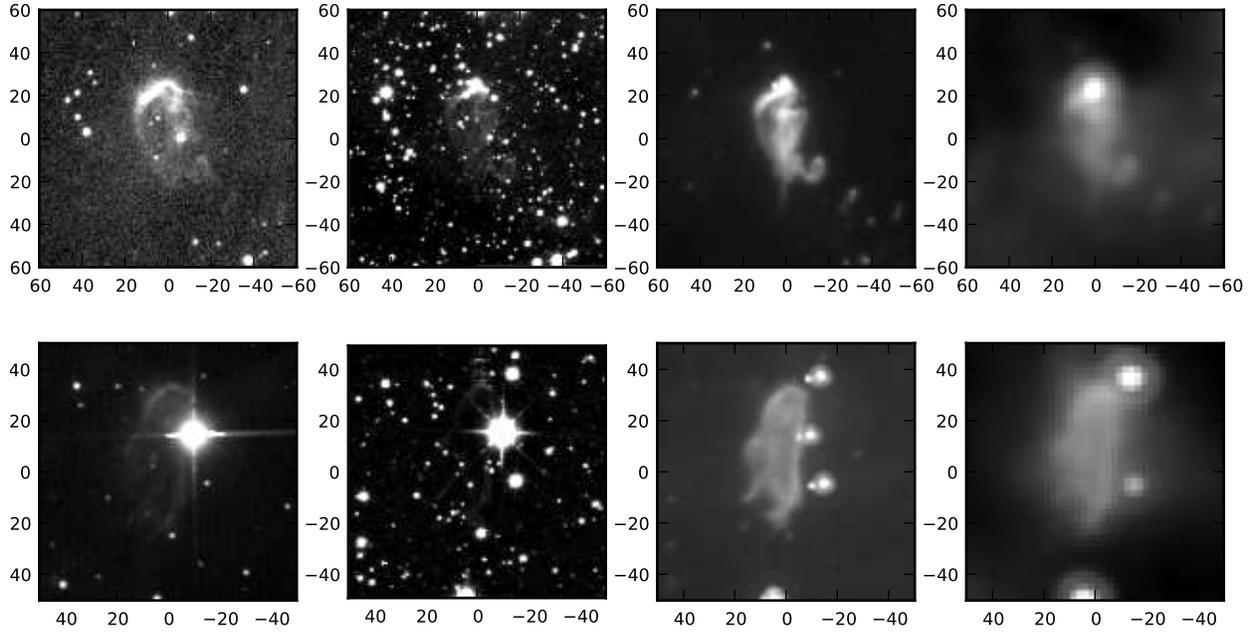}
\caption{Images of objects \#5 (top) and \#6 (bottom) at different wavelengths, from left to right: IPHAS H$\alpha$, UKIDSS K-band, IRAC~8~$\mu$m, MIPS~24~$\mu$m. All images are shown with north up and east to the left and with axes in arcseconds offset from the center of each image. At a distance of 1.40~kpc, $10\arcs \sim 1.4 \times 10^4 \mathrm{AU} \sim 0.07$~pc.}
\label{images2}
\end{figure*}

\clearpage

\begin{figure*}
\includegraphics[scale=0.82, trim=0 -5 -5 -50, clip=true]{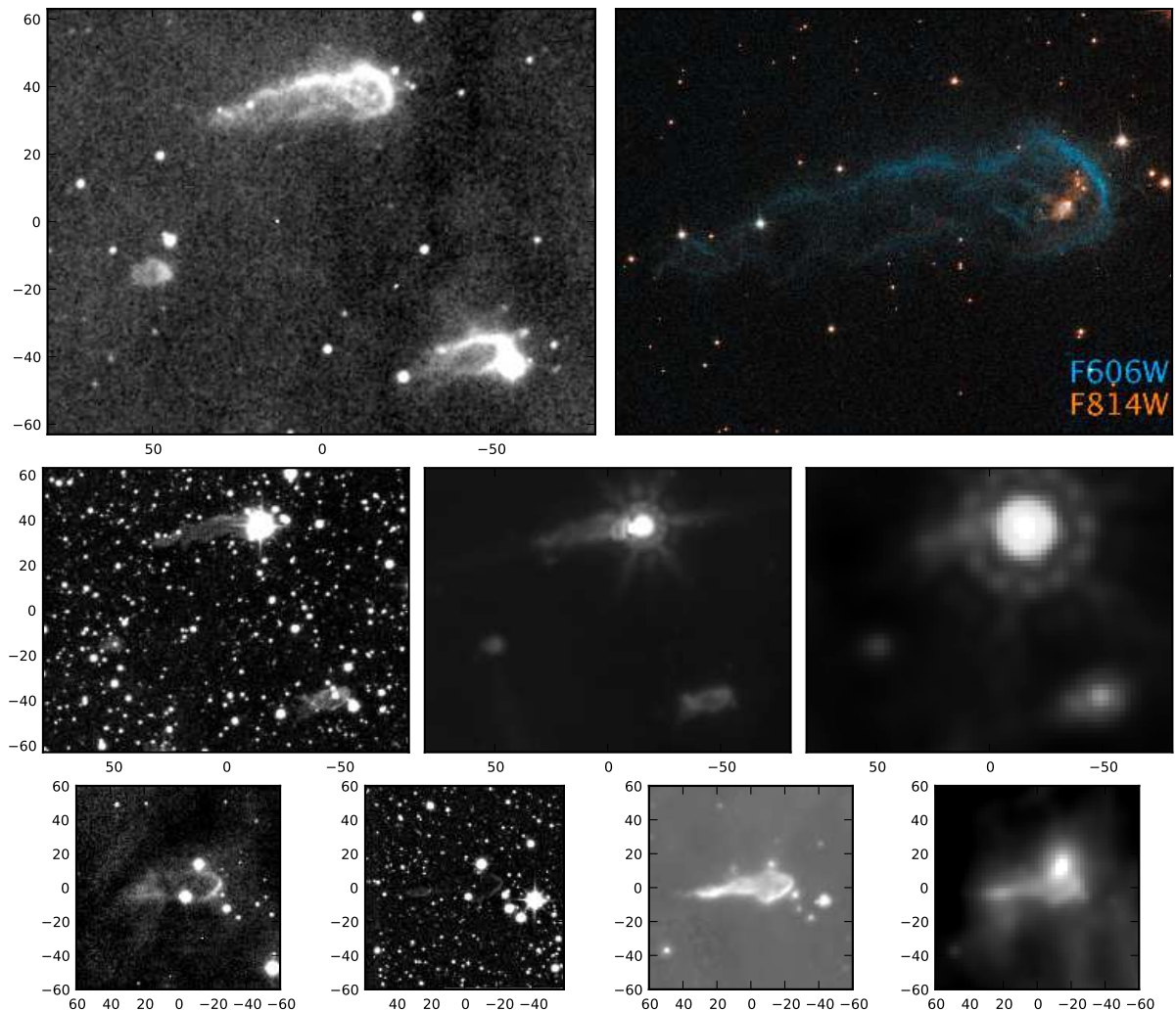}
\caption{Images of objects \#7--10 at different wavelengths. {\it Top left:} Objects \#7--9 in IPHAS H$\alpha$. {\it Top right:} HST/ACS color image of object \#7 (courtesy Z. Levay and L. Frattare, STScI) at a pixel scale of 0.05\arcs ($65\arcs \times 50\arcs \sim 0.44 \times 0.34$~pc). {\it Middle:} Objects \#7--9 in (from left to right) UKIDSS K-band, IRAC 8~$\mu$m, and MIPS 24~$\mu$m. {\it Bottom:} Object \#10, from left to right: IPHAS H$\alpha$, UKIDSS K-band, IRAC~8~$\mu$m, MIPS~24~$\mu$m. All images are shown with north up and east to the left and with axes in arcseconds offset from the centers. At a distance of 1.40~kpc, $10\arcs \sim 1.4 \times 10^4 \mathrm{AU} \sim 0.07$~pc.}
\label{images3}
\end{figure*}

\begin{landscape}
\begin{table*}
\begin{center}
\caption{Properties of the objects identified in the vicinity of Cygnus OB2.}
\label{proplyds}
\begin{tabular}{@{}clccccl}
\hline 
\# & Name & P.A. & Length & Width & Projected & Notes\\ 
 &  & ($^\circ$) & ($10^3$ AU) & ($10^3$ AU) & distance (pc) \\ 
\hline
1 & IPHASX J203453.6+404814 	& 54		& 24 & 16	& 13.7 \\ 
2 & IPHASX J203453.3+405321 	& 55		& 63 & 22	& 12.0 \\ 
3 & IPHASX J203443.2+405313 	& 40		& 113 & 60 & 11.6	& IRAS 20328+4042. Radio source \citep{wend91}.\\ 
4 & IPHASX J203436.4+405154 	& 36		& 102 & 32 & 11.7 \\ 
5 & IPHASX J203311.5+404141 	& 0		& 75 & 41	& 13.9 \\  
6 & IPHASX J203318.8+405905 	& 16		& 77 & 27	& 6.9		& Sub-mm source \citep{roso10}. \\ 
7 & IPHASX J203413.3+410814 	& 80		& 77 & 22	& 5.6		& IRAS 20324+4057.\\  
8 & IPHASX J203410.5+410659 	& 80		& 43 & 15	& 5.8		\\ 
9 & IPHASX J203419.0+410722 	& 85		& 18 & 13	& 6.2\\ 
10 & IPHASX J203447.4+411445 	& 88 		& 83 & 27	& 7.2		& IRAS 20329+4104.\\ 
\hline 
\end{tabular}
\newline 
The source name is provided in the IPHAS sexagesimal, equatorial position-based format for extended objects: IPHASXJhhmmss.s+ddmmss, with ``J'' indicating the position is J2000. The position angle (P.A.) is given as the angle measured clockwise between North and the apparent major-axis of the object. The dimensions of the objects and the projected distance to the center of Cyg~OB2 have been calculated using a distance of 1.40~kpc \citep{hans03}.
\end{center} 
\end{table*} 
\end{landscape}

\end{document}